\documentclass[a4paper]{jpconf}
\usepackage{graphicx}

\usepackage{bm}
\usepackage{amsfonts}
\usepackage{amsmath, amsthm, amssymb}
\usepackage{color}
\newcommand{\gtappr}{{{\lower4pt\hbox{$>$} } \atop \widetilde{ \ \ \ }}}
\newcommand{\ltappr}{{{\lower4pt\hbox{$<$} } \atop \widetilde{ \ \ \ }}}
\newcommand{\beq}{\begin{equation}}
\newcommand{\eeq}{\end{equation}}

\newcommand{\ud}{\mathrm{d}}

\newcommand{\ybal}{$\beta$-YbAlB$_4\,$}

\def\lsim{\buildrel {\textstyle <}\over {_\sim}}

\newsavebox{\fmbox}

\begin{document}
\title{$\boldsymbol{T/B}$ scaling of magnetization in the mixed valent compound \ybal}

\author{Yosuke Matsumoto$^1$, S. Nakatsuji$^1$, K. Kuga$^1$, Y. Karaki$^1$, 
Y. Shimura$^1$, T. Sakakibara$^1$, A. H. Nevidomskyy$^{2, 3}$, P. Coleman$^{2, 4}$}

\address{$^1$ Institute for Solid State Physics, University of Tokyo, Kashiwa 277-8581, Japan}
\address{$^2$ Center for Materials Theory, Department of Physics and Astronomy, Rutgers University, Piscataway, N.J. 08854, USA}
\address{$^3$ Department of Physics and Astronomy, Rice University, Houston, Texas 77005, USA}
\address{$^4$ Department of Physics, Royal Holloway, University of London, Egham, Surrey TW20 0EX, UK}

\ead{matsumoto@issp.u-tokyo.ac.jp, satoru@isso.u-tokyo.ac.jp}

\begin{abstract}
Here we provide the first clear evidence of Fermi-liquid breakdown
in an intermediate valence system. We employ high precision
magnetization measurements of the valence fluctuating superconductor
\ybal to probe the quantum critical free energy down to temperatures
far below the characteristic energy scale of the valence fluctuations.
The observed $T/B$ scaling in the magnetization over three decades not only indicates unconventional quantum criticality, but
places an upper bound on the critical magnetic field $\vert B_c \vert <0.2$~mT, a
value comparable with the Earth's magnetic field and six
orders of magnitude smaller than the valence fluctuation scale. This
tiny value of the upper bound on $B_c$, well inside the superconducting dome, 
raises the fascinating possibility that valence fluctuating \ybal is 
intrinsically quantum critical, without tuning the magnetic field, pressure, or composition: the first known example of such a phenomenon in a metal.
\end{abstract}

\section{Introduction}
The breakdown of Fermi liquid behavior in metals observed near a
magnetic quantum critical point, challenges our current understanding
of strongly correlated electrons. 
%While the mechanism of
%unconventional quantum criticality is actively debated, there is
%a growing consensus that the underlying physics involves a jump in the
%Fermi surface volume, associated with a partial electron
%localization \cite{Schroder00,Coleman01,Si01,Senthil03,Paschen04,Shishido}. 
Archetypical quantum critical (QC) materials can be found in the
heavy-fermion intermetallics, which are
usually described as a Kondo lattice with a competition between
local moment magnetism and conduction electron screening of the local
moments (the Kondo effect) \cite{Mathur98,Lohneysen07,gegenwart08}. 
To date, all QC materials of this kind are known to have an almost
integral valence which stabilizes the local moments considered
essential for the criticality. By contrast, departures from integral
valence associated with valence fluctuations are thought to promote
screening of local moments, suppressing critical phenomena.

On the other hand, recent studies by our group revealed the first example of an 
Yb-based heavy fermion superconductivity in the new compound \ybal ($T_c$ = 80 mK) \cite{nakatsuji08,KugaPRL}. 
Pronounced non-Fermi liquid behavior above $T_c$ and its magnetic field dependence indicate that the system is a rare example of a pure metal 
that displays quantum criticality at ambient pressure and close to zero magnetic field \cite{nakatsuji08}. 
Furthermore, recent hard x-ray photoemission spectroscopy measurements revealed 
intermediate valence of Yb$^{+2.75}$ providing the first unique example of quantum criticality in a mixed valent system \cite{ybal-valency}. 
Interestingly, 
the system is governed by two distinct energy scales: a high-energy scale, $T_0 \sim 200$~K
characterizing valence fluctuations, and a low-energy scale, 
$T^*\sim 8$~K characterizing the emergence of the Kondo lattice behavior~\cite{matsumoto-ZFQCP}.

Here we present clear experimental evidence that indicates the zero-field quantum criticality in \ybal \cite{matsumoto-ZFQCP}. 
We employed high resolution magnetization measurements and found a scaling equation $- dM/dT = B^{-1/2}f(T/B)$, 
which holds over wide ranges of magnetic fields (0.3 mT $< B <$ 2 T) and temperatures (20 mK $< T <$ 3 K). This $T/B$ scaling strongly suggests $B_c$ = 0 
within the experimental resolution of $\sim$ 0.2 mT, 
which is comparable to the Earthfs magnetic field. 
This indicates a possibility of quantum critical phase that extends over a finite pressure range.
The experimental details of this work have been discussed already in~\cite{matsumoto-ZFQCP,matsumoto_PSSB247}.

\section{Results and discussion} 
The divergent magnetic susceptibility $\chi$ at $T \to 0$ is one of the most remarkable feature 
which characterize the QC in \ybal.
 Figure 1 shows the $T$ dependence of the $c$-axis susceptibility $\chi = M/B$ in the wide range of $T$ and $B$ 
spanning four orders of magnitude. 
Below $T\sim$ 3 K, 
the results show a systematic evolution 
from a non-Fermi liquid (NFL) metal with divergent susceptibility at zero field to a Fermi liquid (FL) 
with finite susceptibility in a field.

\begin{figure}[bt]
\begin{center}
\includegraphics[width=36pc]{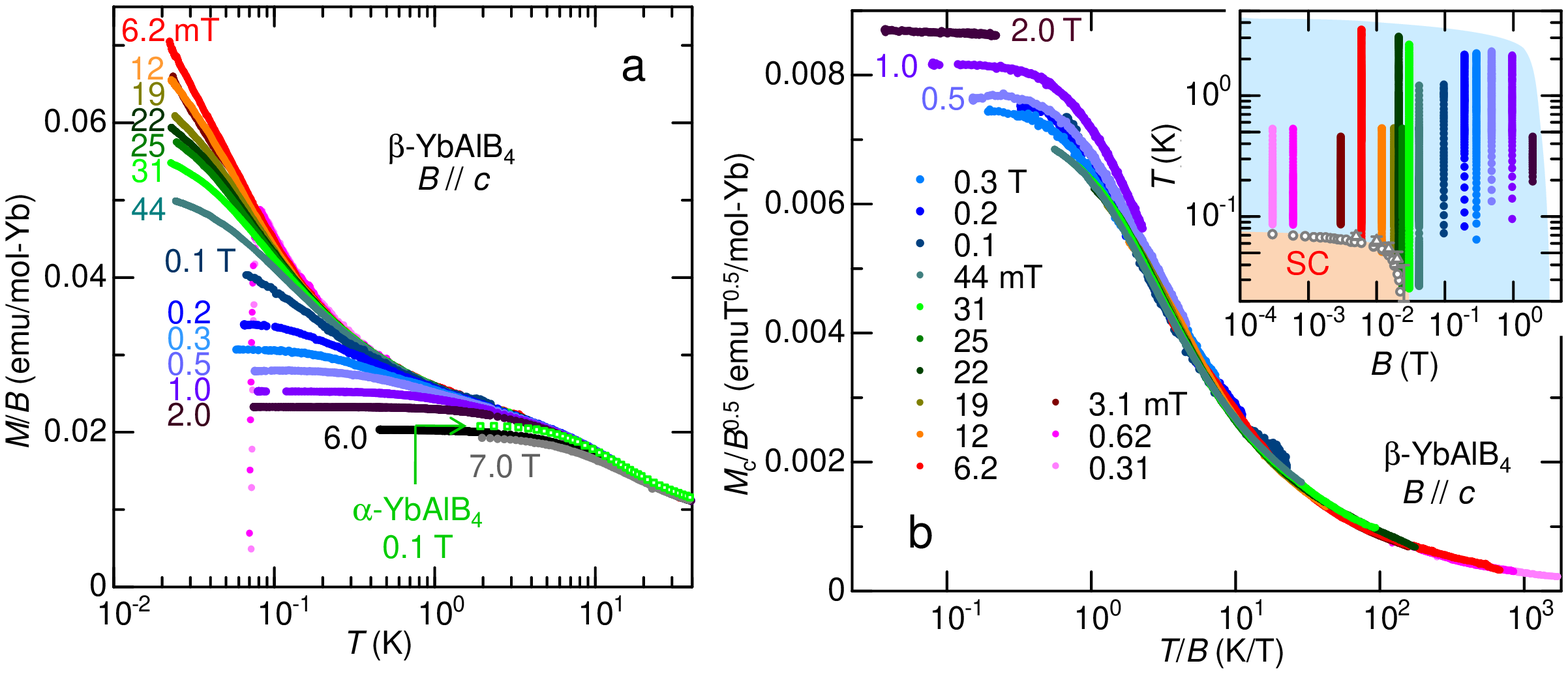}
\end{center}
\caption{\label{magnetization}(a) Temperature dependence of the magnetic susceptibility 
$M/B$ of both $\beta$- (solid circles) and $\alpha$-YbAlB$_4$ (open squares) under the field $B\parallel c$. 
(b) Scaling plot of $M_c/B^{0.5}$ vs. $T/B$ for \ybal. 
$M_c$ is the critical contribution to the magnetization: \mbox{$M_c = M - M_0$}(see text). 
Inset shows the $B$-$T$ range where the scaling applies (solid
circles in the blue shaded region) and the superconducting (SC) phase boundary (open circles and triangles~\cite{KugaPRL,matsumoto_PSSB247}) of \ybal.
}
\end{figure}

Divergent behavior in NFL region is well fitted by a sum of the QC contribution $M_c/B\propto T^{-0.5}$ and 
$T$-independent constant term $\chi_0 (= M_0/B) = 0.017$ (emu/mol). 
The latter constant term, which is close to the zero-$T$ susceptibility of the non-critical $\alpha$-YbAlB$_4$, 
may originate from constant Van Vleck contribution to susceptibility and/or from Pauli susceptibility of the non-critical parts of the Fermi surfaces. 
Figure 1b shows $M_c/B^{0.5}$ vs. $T/B$ for \ybal in
the region shown in the inset ($T \lesssim 3$~K and $B \lesssim 2$~T). 
Interestingly, 
the data collapse on a single curve at $T>B$, indicating a scaling relation $M_c =M - M_0 = B^{0.5}\psi (T/B)$. 
Under fields $B > 0.5$ T, $M_{\rm c}$ does not follow a single scaling curve. This may be due to a small field-nonlinear contribution to $M_0$, which is neglected in the above analysis. 
In fact, the scaling relation free from this contribution which is obtained after $T$-derivative of the both parts: 
\beq 
-\frac{\ud M}{\ud T} = B^{\alpha -2}\phi \left(\frac{T}{B} \right),  
\label{dMdT} 
\eeq 
with $\alpha = 3/2$ works much better as shown in Figure 2a. %Here $\phi (x) = \psi '(x)$.
Here we can observe a clear power law behavior of $-dM/dT\propto T^{-1.5}$ at $T>B$ (NFL region), which indicates  
$M\propto T^{-0.5} + const.$, and $T$-linear behavior of 
$-dM/dT$ at $T < B$ as expected for a FL. 
This empirical scaling property 
implies that close to the QCP, below $T\sim3$~K and
$B\sim 2$~T, \ybal has no intrinsic energy scale, and that furthermore,
temperature and field are interchangeable variables, whose
ratio $T/B$ determines the physical properties. 
The peak of the scaling curve is located at $T/B \sim 1$ K/T and indicates that the thermodynamic boundary between the FL and NFL regions is on the $T \sim B$ line, as shown in the phase diagram of the inset of Figure 2a. 
The QC free energy with the form $F_\text{QC}= B^{\alpha} f(T/B)$ is obtained by integrating both parts of Eq.~(\ref{dMdT}), 
where $f$ is a scaling function of the ratio $T/B$ with the limiting behavior:
$f(x)\propto x^{\alpha}$ in the NFL regime ($x \gg 1$) and $f(x)\propto \mathrm{const} + x^2$ in the FL phase ($x\ll 1$). Indeed, $f(x) \propto (A+x^2)^{\alpha/2}$ fits the observed scaling behavior of $dM/dT$ in Eq. (1) with $\phi(x) = \Lambda x(A+x^2)^{\frac{\alpha}{2}-2}$, 
as shown by the fit in Figure 2a, achieved with $\alpha=3/2$.
%From this function $\phi$, one can obtain a very simple form of the free energy:
%$F_\text{QC} =  -(\kB \tilde{T})^{-1/2}((g\mu_\text{B}B)^2 + (\kB T)^2 )^{3/4}$, 
%which gives the best fit with the effective moment $g\mu_\text{B} = 1.94 \mu_B$ and the energy scale $\kB\tilde{T}\approx6.6$~eV. 

\begin{figure}[bt]
\includegraphics[width=19pc]{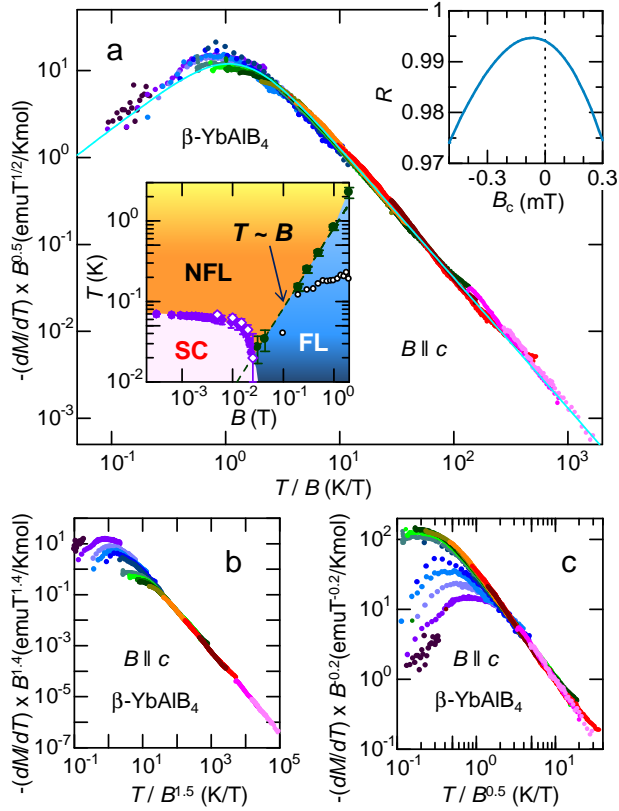}\hspace{2pc}%
\begin{minipage}[b]{16pc}\caption{\label{scaling}(a) Scaling observed for the $c$-axis magnetization at $T \lsim 3$ K and $B \lsim 2$ T. 
The light blue line represents a fit to the empirical Eq. 1 with scaling function $\phi(x) = \Lambda x (A+x^2)^{-n}$
%, 
%where $x = T/B$ 
(see text). 
%The best fit is obtained with $n = 1.25 \pm 0.01$ and $B_c = -0.1 \pm 0.1$ mT (light blue line), corresponding to $\alpha=3/2$ in the scaling form of the 
%free energy, Eq.~2, see \cite{MM}, and $\vert B_c \vert< 0.2$ mT. 
The right inset shows Pearson's correlation coefficient $R$ for the fit with finite $B_c$ (see text). 
%$R$ reaches a maximum value of 1 if the fit quality is perfect. 
The best fit is obtained with $\alpha = 1.50 \pm 0.02$ 
and $B_c = -0.1 \pm 0.1$ mT (see text).
The left inset shows the $B$-$T$ phase diagram of \ybal in the low $T$ and $B$ region. 
The filled circles are the peak temperatures of $ -dM/dT$ below which the FL ground state is stabilized. 
%At %low field, the thermodynamic boundary between the FL and NFL regions is on a $k_{B}T \sim g\mu_{\rm B}B$ line (broken line). 
The open circles are the temperature scale $T_{\rm FL}$ below which the $T^2$ dependence of the resistivity is observed~\cite{nakatsuji08}.
The upper critical fields of the superconducting (SC) phase is shown by open circles and triangles~\cite{KugaPRL,matsumoto_PSSB247}. 
(b)(c) Scaling plots with ($\delta$, $\eta$) = (1.5, 1.4), (0.5, -0.2) for the same data set in Fig.~2a, respectively (see texts). }
\end{minipage}
\end{figure}

The $T/B$ scaling implies that the critical field $B_c$ of the quantum
phase transition is located just at zero field. A finite value of $B_c$ requires the scaling function 
$f(x)$ and $\phi(x)$ with a ratio $x=T/|B-B_c|$ rather than $x=T/B$. 
A bound for $B_c$ can be determined by substitution of $x=T/|B-B_c|$ into Eq.~(\ref{dMdT}) while searching the value of $B_c$ 
that gives the best fit to the experimental data. The inset of Figure 2a shows the Pearson's correlation coefficient $R$ obtained from this fit, 
indicating that $B_{\rm c}$ is optimal at $-0.1 \pm 0.1$~mT. 
The error-bar is only a few times larger than
Earth's magnetic field ($\sim 0.05$~mT), more significantly it is two orders of magnitude smaller than $\mu_0 H_{\rm c2} = 30$ mT. This result is
particularly stunning given the large valence fluctuation scale $T_0\sim$200~K.
Thus \ybal provides a unique example of essentially \emph{zero-field quantum criticality}.

A possibility of other scaling, such as $T/B^{\delta}$ scaling ($-\ud M/\ud T = B^{-\eta}\phi (T/B^{\delta})$) with $\delta \neq 1$, 
is checked by scaling plot, $-(\ud M/\ud T)B^{\eta}$ versus $T/B^{\delta}$. 
Figures 2b and 2c show the trial scaling plots with ($\delta$, $\eta$) = (1.5, 1.4), (0.5, -0.2) for the same data set in Fig.~2a, respectively. 
Here, $\eta$ was adjusted against fixed $\delta$ so that a scaling was realized at high $T$ parts where the NFL power-law behavior was observed. 
As is clearly seen from the figures, 
the data at smaller $T/B^{\delta}$ do not satisfy the scaling by changing $\delta$ away from 1. 
It turned out that 
the data in the widest $T$ and $B$ range are scaled with $\delta = 1.0 \pm 0.1$.  
The detail will be discussed elsewhere~\cite{matsumoto_full}. 

%Furthermore, as shown in the right panel of Fig.~4, $\Gamma_H/H$ show a systematic change in the temperature dependence, from high-field nearly constant behavior, which is expected for FL state, to a strongly temperature dependent one under low field. Particularly, in the low field limit, or the NFL regime, $\Gamma_H/H$ diverges on cooling. The best fit to the data taken under 0.31 mT shows a strong power-law temperature dependence $\Gamma_H\sim B/T^{2.4}$. 
%This divergence of $\Gamma_H/H$ at $B \to 0$ provides another evidence for the zero-field quantum criticality. 
%The apparently large power law exponent of $\Gamma_H$ can be consistently understood within the phenomenological scaling of the free energy Eq.~(\ref{free-energy}) (Supplementary Information).
%, from which one obtains $\frac{\ud M}{\ud T} \sim T^{-3/2}$, and assuming roughly constant specific heat coefficient $\gamma=C_v/T$, disregarding a weak log-dependence in Eq. (\ref{Cv-fit}), one immediately obtains $\Gamma_H\sim B/T^{2.5}$ from Eq.~(\ref{Gamma_H}), in good agreement with experiment.

The observation of zero field quantum criticality in valence fluctuating \ybal has a number of significant implications. Conservatively, 
one might conclude that \ybal is a system in which a fortuitous combination of
structure and chemistry fine-tune the
critical field  $B_{c}$ to zero. 
In such a scenario, the QCP would be characterized by gapless
soft modes:  spin fluctuations in the case of an antiferromagnetic quantum criticality, or critical charge fluctuations associated with mixed valence. In the former case, the system would be literally at the edge
of an antiferromagnetic instability, 
while  the latter case would involve a 
critical end-point of valence instability line~\cite{Yuan03, Holmes04}.
% analogous to a liquid-gas transition at $T=0$,
%and similar to the QC end-point observed in Sr$_3$Ru$_2$O$_7$~\cite{SrRuO-327-endpoint}. 
In both cases, the application of pressure is expected to immediately restore the Fermi liquid ground state.

Set against these two possibilities is our observation of a
tiny upper bound (0.2~mT) on the critical magnetic field $B_c$. 
This critical state is so fragile, 
that a field not much larger than the Earth's magnetic field drives \ybal 
back into a Fermi liquid. 
This suggests the intriguing possibility that  
\ybal is a \emph{quantum critical
phase} of matter, to which infinitesimal magnetic field is a relevant
perturbation converting it into a Fermi liquid. In this case, non-Fermi liquid properties will be unaffected by pressure.  An example of such a state is a gapless spin liquid, %\cite{gapless-SL}, 
in which the presence of an extensive manifold of gapless spin excitations would lead to a critical phase 
characterized by power-law correlation functions~\cite{Senthil03,Paul07}. 
It is left for future studies, such as pressure experiments, to discriminate 
between these two alternative scenarios.

%\begin{figure}[t]
%\begin{minipage}[t]{18pc}
%\includegraphics[width=18pc, clip]{fig1.eps}
%\caption{\label{magnetization}Temperature dependence of the magnetic susceptibility 
%$M/B$ of both $\beta$- (solid circles) and $\alpha$-YbAlB$_4$ (open squares). 
%The Curie-Weiss fit above 150 K yields a Weiss temperature $\Theta_{\rm W} \sim W%-110$ K and an effective moment of $\sim 2.2 \mu _B$ for both systems.
%Inset shows $-dM/dT$ versus $T$ for selected fields on a logarithmic scale for both $\beta$- (solid circle) and $\alpha$-YbAlB$_4$ (open square).
%Below $T\sim$ 3 K, $-dM/dT$ for the $\beta$ phase shows $T^{-1.5}$ dependence  
%(broken line) in the non-Fermi liquid (NFL) region at $T > B$, and $T$-linear behavior (solid line) as expected for a Fermi liquid (FL) at $T < B$. }
%\end{minipage}\hspace{2pc}%
%\begin{minipage}[t]{18pc}
%\includegraphics[width=18pc, clip]{fig1.eps}
%\caption{\label{HoverM}Temperature dependence of inverse susceptibility $H/M$
%in $B=$ 0.1 T along the $c$ axis. Solid and broken lines are Curie-Weiss fits above 150 K for 
%$\alpha$-(green) and \ybal (red symbols), respectively. 
%Inset shows low temperature part of $H/M$. Lines are Curie-Weiss fits below 20 K (see text). 
%}
%\end{minipage} 
%\end{figure}

\ack
We thank H. Ishimoto, D.E. MacLaughlin, K. Miyake, T. Senthil, 
Q. Si, T. Tomita, K. Ueda, and S. Watanabe for useful discussions.
This work is partially supported by Grants-in-Aid (No.
21684019) from JSPS, by
Grants-in-Aids for Scientific Research on Priority Areas (No. 19052003) and
on Innovative Areas (No. 20102006, No. 20102007) from MEXT, Japan, by Global
COE Program ``the Physical Sciences Frontier", MEXT, Japan, by Toray Science
and Technology Grant, and by a grant from the
National Science Foundation DMR-NSF-0907179 (P. C. and A. H. N.). 
%P. C. and A. H. N. acknowledge the hospitality of the Aspen Physics Center.

\section*{References}
\providecommand{\newblock}{}

%\bibliography{matsumoto_SCES2011}

\end{document}